# Peculiarities of the transport properties of InMnAs layers, produced by the laser deposition, in strong magnetic fields

V.V. Rylkov, A.S. Lagutin, B.A. Aronzon
*Russian Research Center «Kurchatov Institute», 123182 Moscow, Russia*

V.V. Podolskii, V.P. Lesnikov
*Physico-technical institute of the Nizhniy Novgorod State University,
603950, Nizhni Novgorod, Russia*

M. Goiran, J. Galibert, B. Raquet and J. Leotin
*Laboratoire National des Champs Magnétiques Pulsés, 143, Avenue de Rangueil-BP 14245,
31400 Toulouse Cedex 4, France*

**Abstract**

Magnetotransport properties of p-InMnAs layers are studied in pulsed magnetic fields up to 30 T. Samples were prepared by the laser deposition and annealed by ruby laser pulses. Well annealed samples show p-type conductivity while they were n-type before the annealing. Surprisingly the anomalous Hall effect resistance in paramagnetic state ($T > 40$ K) and in strong magnetic fields ($B > 20$ T) appears to be greater than that in ferromagnetic state ($T \leq 40$ K), while the longitudinal resistance rises with the temperature decrease. The negative magnetoresistance saturates in magnetic fields higher then 10 T at $T \approx 4$ K only, whereas the saturation fields of the anomalous Hall effect resistance are much less ($\approx 2$ T at $T \approx 30$ K). The total reduction of resistance exceeds 10 times in magnetic fields around of 10 T. The obtained results are interpreted on the base of the assumptions of the non-uniform distribution of Mn atoms acting as acceptors, the local ferromagnetic transition and the percolation-like character of the film conductivity, which prevailed under conditions of the strong fluctuations of the exchange interaction. Characteristic scales of the magneto-electric nonuniformity are estimated using analysis of the mesoscopic fluctuations of the non-diagonal components of the magnetoresistivity tensor.

## INTRODUCTION

The interest for studies of diluted magnetic semiconductors (DMS) of the type III-Mn-V, has sharply increased after the discovery of ferromagnetism in In1-xMnxAs films with the Curie temperature $T_c \approx 7.5$ K [1,2]. These magnetic semiconductors can easily be grown on single crystal GaAs substrates that opens prospects to develop new spintronics devices [2,3]. Studies of $In_{1-x}Mn_xAs$ are still actual despite of the Curie temperature doesn't exceed 50 K [4] being considerably less than that in $Ga_{1-x}Mn_xAs$ ($T_c$ = 159 K [5]). In particular, this urgency is caused by the possibility to control the ferromagnetic ordering in the given material using the field effect [6] or the illumination exciting of the non-equilibrium carriers [7]. Magnetic properties of $In_{1-x}Mn_xAs$ films are strongly correlated with the type of conductivity. Samples with n-type conductivity are paramagnetic, while p-type samples show the anomalous Hall effect (AHE) [2]. On the other hand the disorder and peculiar structural properties of the material to a grate extent determine it's magnetic state. For example, In1-xMnxAs films prepared by the low-temperature molecular beam epitaxy (LT-MBE) could be both n- or p-type depending on the growth temperature and the kind of substrate [2]. Alloying of InMnAs with GaMnAs leads to the substantial increase of $T_c$ (up to 110 K) obtained at the hole concentration about one order of magnitude smaller than in case of $Ga_{1-x}Mn_xAs$ samples. Possibly, that is due to the short-range order effects in the Mn atoms arrangement and with the formation of dimers [4,8]. Those effects are assumed to be responsible for observation of the ferromagnetic state at temperatures up to $T$ = 333 K in $In_{1-x}Mn_xAs$ films, obtained by the metal-organic vapor phase epitaxy (MOVPE) [9]. However in those samples ferromagnetic state weakly affects the transport properties and they have a small hole concentration ($10^3$ times less compared with the Mn concentration) [10].

Recently we have shown that films with high Mn concentration (≥10 at.%) in the III-V semiconductors can be successfully obtained by the deposition from the laser plasma (DLP) in vacuum [11,12]. The GaMnSb films fabricated by this method showed AHE with a hysteresis loop at temperatures up to the room that and their coercivity growing up with increasing hole concentration [11]. The InMnAs films, obtained in similar manner, have the n-type conductivity and are paramagnetic. However, it was found that the inversion to the p-type could be reached after annealing by ruby laser pulses (the pulse duration is about of 25 nanoseconds; the pulse energy is about of 1 J). As a result, the AHE appears at liquid nitrogen temperature testifying the ferromagnetic state of the film [12].

In this paper we present the results of magnetotransport studies of p-InMnAs layers, in magnetic fields up to 30 T. Additionally to observation ferromagnetic state at $T \leq 40$ K we surprisingly observed AHE under the hopping regime of conductivity, unsaturated negative magnetoresistance and some other peculiar features. Furthermore AHE resistance in paramagnetic state was found to be larger than that at temperatures lower the Curie value. The obtained results are interpreted taking into account the non-uniform distribution of Mn acceptors, the local ferromagnetic transition and the percolation-like character of the film conductivity, which are prevailing under the conditions of strong fluctuations of the exchange interaction.

**EXREIMENTAL AND RESULTS**

The $In_{1-x}Mn_xAs$ single crystalline films of a mosaic type with the Mn contents about of 10 at.% and the thickness about of 200 nm have been grown on GaAs semi-insulating substrates by means of the DLP technology [11,12]. The temperature of the film growth was about 200 °C. In contrast with GaMnSb, InMnSb and GaMnAs films, neither the magneto-optical Kerr effect nor the ferromagnetic resonance absorption were detected in case of InMnAs [12,13] that specified the absence of the ferromagnetic inclusions of MnAs. Several samples were measured showing analogous behavior but results presented in this paper are related to the sample which has longitudinal resistivity $\rho_{xx} \approx 3 \cdot 10^{-2}$ Ω·cm at $T = 300$ K and the hole concentration $p \approx 2.6 \cdot 10^{19}$ cm$^{-3}$ obtained from the Hall effect measurements at fields $B < 1$ T.

Samples for the magnetoresistance and the Hall effect measurements were of the standard double cross shape (the width and length of the conduction channel were $W = 2.5$ mm and $L = 9$ mm, respectively). Measurements were carried out in the temperature range from 4 to 100 K in pulsed magnetic field up to 30 T; the pulse duration was equal 0.8 s and the field rising time was around of 0.08 s. Magnetotransport characteristics were analyzed for two field polarities and during falling edge of magnetic field only.

Figure 1 shows the temperature dependence of the specific resistance ρxx of the sample[1]. Such temperature dependence is similar to that observed in ferromagnetic DMS and, particularly, in $Ga_{1-x}Mn_xAs$ films [2,3,14-16] with an activation type of conductivity ($\rho_{xx} \geq 2 \cdot 10^{-2}$ Ω·cm [14,15]). This type of the temperature dependence corresponds to the well known maximum of ρxx(T) observed in ferromagnetic materials close to Curie temperature. From the Fig.1 one can see that the resistance grows noticeably when the temperature decreases approaching 40 K. However this temperature, the tendency to plateau (a local minimum of $\rho_{xx}$(T)) is observed followed by the abrupt (activation-like) growth of the resistance. The corresponding peculiarities are marked by $T_c$ at Fig. 1. The position of this local maximum of $\rho_{xx}(T)$ is frequently used to determine the Curie temperature [3,15]. In our case, such estimation gives the Curie temperature $T_c \approx 40$ K (see Fig. 1).

---

[1] These measurements of $\rho_{xx}$ were carried out in steady magnetic fields about 100 Oe, caused by the residual current passing through the coil.

Except of the $T_c$ vicinity the temperature dependence of resistivity is of the activation type. The best fitting of $\rho_{xx}(T)$ curve could be obtained by the expression $\ln\rho_{xx} \propto (T_0/T)^{1/4}$ (see insert in Fig. 1) that corresponds to the Mott mechanism of the variable-range hopping conductivity [17]. Note that similar mechanism is realized in compensated $Ga_{1-x}Mn_xAs$ samples with the insulating type of the conductivity [15,16]. However at $T<T_c$ the experimental results slightly deviate from the Mott law.

Magnetic field dependencies of the Hall resistance $R_H(B)$ at temperatures lower ($T \leq T_c \approx 40$ K, curves 1 and 2) and above Curie that (curve 3, $T = 88$ K) are shown in Fig.2. Low temperature curves are typical for the anomalous Hall in ferromagnetic state and $R_H(B)$ at $T = 88$ K does not look unusual for AHE in a paramagnetic material. In both cases curves are similar to those observed in $Ga_{1-x}Mn_xAs$ layers with the metallic type of conductivity under conditions of substantial domination of the anomalous Hall component over normal one [2,14]. However, contrary to the ordinary situation and particularly to $Ga_{1-x}Mn_xAs$ samples, the AHE resistance of InMnAs sample under paramagnetic regime (curve 3), in fields above 20 T exceeds the saturation resistance of the AHE (RAS) of that sample in ferromagnetic state (at $T \leq T_c \approx 40$ K). That is surprising. It is necessary to remind that in single-phase III-Mn-V DMS, as well as in ferromagnetic metals, the Hall resistance $R_H$ follows to relation [3]:

$$R_H d = \rho_{xy} = R_0 B + R_s M \qquad (1)$$

where $d$ is the sample thickness, $R_0$ is the normal Hall effect coefficient and $R_s$ is the AHE coefficient. The first term is due to the Lorentz force and the Hall resistance in this case is proportional to the magnetic induction $B$, while anomalous Hall effect is proportional to the magnetization M and is due to spin–orbit interaction. There is a few known mechanisms for AHE and according all of them AHE in DMS depends on the strength of the spin-orbit interaction as well as on the spin polarization of carriers and follows the relation $R_s \propto (\rho_{xx})^\alpha$. For the "skew scattering" mechanism $\alpha = 1$ and $\alpha = 2$ for the "internal" and "side jump scattering" mechanisms [3].

In our case the AHE contribution is dominating (see Fig. 2) and $R_H \approx (R_s/d) \cdot M$. So the fact that the $R_H$ at $T = 88$ K is higher than that at $T \leq T_c$ looks like the saturated magnetization $M_s$ in paramagnetic state is larger than the value of $M_s$ in ferromagnetic state in the same sample. Let us note that it is impossible to explain this very surprising result in the frame of any of known mechanisms of the AHE [3] if we do not make any assumption of sample structure. Really $R_s \propto (\rho_{xx})^\alpha$ and $\rho_{xx}$ grows with the temperature decrease being at $T \approx 40$ K almost 3 times greater than at $T = 88$ K. This peculiarity is not related with presence of the normal Hall effect because its contribution is very small (see slope of the curve 1 in Fig.2). Note also, that the hole concentration found from $R_0$ at $T$ = 25 K equals $p \approx 4.8 \cdot 10^{19}$ см$^{-3}$ and exceeds the room temperature value less than two times.

Also it should be noted that the residual AHE resistance (at $B$ = 0 T and $T$ = 25 K) achieves high enough value 20 Ω, approaching about of 40% of the saturation AHE resistance RAS, as it is seen in the insert to Fig. 2. That testifies the presence of the residual magnetization of the sample and its ferromagnetic ordering.

Along with specific behavior of the AHE in InMnAs layers the magnetoresistance of those structures was also studied and results are presented at Fig. 3. The negative magnetoresistance (NMR) was observed and it's value rises when temperature diminishes approaching the ferromagnetic transition. Such behavior is known for III-Mn-V DMS [1,2,14]. At temperatures lower than the Curie that on the further temperature decrease the NMR value either falls down [2,14] (for samples with the metal type of conductivity) or NMR continues to increase. The latter was observed just in our measurements (see Fig. 3). NMR appears under the paramagnetic regime, grows in the vicinity of the ferromagnetic transition and monotonously increases with temperature decrease up to $T$ = 4.2 K. Under these conditions resistance $R_{xx}$ decreases more than by 10 times. Previously, for InMnAs films with activation type of conductivity the analogous observation was made by H. Ohno et al [1] discovered very large drop of $R_{xx}$ (by about 5 times) at $B \approx 10$ T and $T$ = 2.8 K. This huge NMR was qualitatively explained there by the lowering of the activation energy between the local magnetic polaron state and the delocalized state above the mobility edge.

The following should be stressed, there is no even the tendency of resistance to saturate with rising magnetic field even in fields up to $B \approx 30$ T, while AHE, determined by the sample magnetization, achieves the saturation at field $B \approx 2$ T (see insert in Fig. 2). Only at $T$ = 4.2 K and at $B \approx 20$ T there is some features resembling such tendency (see insert in Fig. 3). Basically, the absence of NMR saturation in high magnetic fields was observed in GaMnAs with the metallic type of conductivity, where it was attributed to the quantum corrections to conductivity [18]. However, it is not the case for the results under discussion, because the conductivity increase is large enough, and the sample conductivity is of the hopping type, hence the quantum corrections to conductivity contribution could not dominate. It should be noticed also that the observable change of resistance exceeds essentially the MR magnitude for tunnel structures "ferromagnet/dielectric/ferromagnet", where conductivity increases less than by 2 times (see, for example, [19]).

Thus the following observations should be noted and explained: the local maximum on the temperature dependence of resistivity, which is the sign of the ferromagnetic transition and the Curie temperature; both normal and anomalous Hall effect under regime of the hopping conductivity; the surprising exceeding of AHE at paramagnetic regime compared to its saturation value at the ferromagnetic state; and the large NMR value, which does not saturate in fields up to 30 T contrary to AHE, which saturates at 2 T.

**DISCUSSION**

First of all let points out, that the Hall effect is extremely small under the hopping regime [20], and there are no experimental data concerning its behavior in conventional semiconductors. The anomalous Hall effect is stronger. It was observed in Fe/SiO2 nanocomposites in the hopping regime, however in the direct vicinity of the percolation transition and at temperatures $T > 77$ K [21]. So it is difficult to explain the above results within the framework of the model of the homogeneously doped semiconductor.

The situation becomes clear, if one takes into account the structural features of InMnAs and assumes that after the pulsed laser annealing the acceptor atoms of manganese are nonuniformly distributed in the sample. It was already mentioned that before annealing InMnAs samples are of n-type and only annealing activates part of Mn atoms to act as acceptors. This statement is also valid for InMnAs, produced by the LT-MBE method [2]. That is due to high concentration of defects and also owing to short-range effects in Mn distribution (formation of dimers, for example), which suppress by orders of magnitude its electrical activity at high Mn content ($x \approx 0.1$) [4,8-10]. We have found that the pulsed laser annealing promotes the substantial activation of the Mn impurities (the increase of the holes concentration) not only in case of InMnAs, but also for GaMnAs and GaMnSb films [12,13]. However, apparently, this nonequilibrium process leads to the non-uniform distribution of Mn acceptors causing the appearance of areas both with the increased hole concentration (degenerated or "metallic droplets") and with reduced that (strongly compensated region). It could be seen in Fig. 4 which shows the schematic representation of a bending of valence band top caused by nonuniform Mn acceptors distribution and forming the hole droplets. With temperature decrease the local ferromagnetic transition inside the droplet results in lowering of the hole energy and so makes potential wells deeper. Under this assumption all observed peculiarities could be naturally explained.

Indeed, according this picture the sample resistance should be determined by the most resistive compensated regions, and because the hopping conductivity starts to dominate from the room temperature, the Fermi's level in these regions is located below the valence band top at the energy considerably exceeding the thermal that. In this situation the Hall effect is natural to describe in the framework of the percolation two-component systems with different conductivities like metal - bad conductor [22]. Droplets with the high hole concentration acts as a metal, and the areas between them correspond to bad conductor ("dielectric"). Following [22] one can write:

,

$$R_H \approx R_m + R_d \frac{\sigma_d^2}{\sigma^2} \tag{2}$$

where $R_m$, $R_d$ are the Hall resistance in drops and dielectric spacers, and $\sigma$, $\sigma_d$ are the effective conductivity of medium and dielectric spacers, accordingly. Value of $R_d$ equals ≈0 for the hopping conductivity [20,21]. Therefore, in our case, the Hall resistance is determined substantially by the hole drops, where the effective concentration of carriers $p$ is 2.6·10$^{19}$ см$^{-3}$ in accordance with results of the Hall effect measurements at $T$ = 300 K. Following ref. [4], at such concentrations the ferromagnetic transition occurs in homogeneously d`ped In$_x$Mn$_{1-x}$As films (x≈0.1) in the temperature range from 30 K to 50 K, that is in accordance with previously mentioned value $T_c$ ≈40 K in our case. This transition, however, should have a local character in studied samples, taking place inside the hole drops only. That is because the magnetic interactions in DMS is carrier mediated as commonly accepted [3].

For this description and the above model to be valid the drop size should be big enough, it should be longer mean free path and length of magnetic interaction. That is needed for possibility to describe the conductivity and magnetic ordering inside the drop in terms of classical carrier scattering and magnetic transition. So let estimate the droplet size.

The residual RH being the sign of residual magnetization appears at $T ≈ 40$ K. So drops magnetization are blocked, that means the energy of magnetic anisotropy Va exceeds the thermal energy by more then 20 times. Taking into account that the energy of magnetic anisotropy connected with the shape of drops only [23] and assuming that their saturation magnetization is about of 100 G [4], one could estimates the drop sizes to be above 20 nm. It should be larger than the mean free path for holes in III-Mn-V DMS (its typical value is about 0.6-0.8 nm in GaMnAs [24]).

If magnetic particles are isolated from each other and axes of their magnetic anisotropy are arranged randomly, the residual magnetization Mr of a sample at zero magnetic fields equals $M_r=(1/2)M_s$ [25] (here $M_s$ is the saturated magnetization). That is in an agreement with AHE measurements (see Fig. 2) which give $M_r ≈ 0.4·M_s$. So it looks like magnetic metal-insulator nanocomposites at the insulating side of the percolation transition. In our case, however, the conductivity is due to the hopping through the impurity band in compensated areas of InMnAs sample but not by the direct tunneling between metal drops, as it occurs in nanocomposites in the vicinity of percolation threshold. That could be proved by the temperature dependence of conductivity in measured films which follows the Mott low but not by the law "1/2" (ln$\rho_{xx}$ ∝

*[T₀/T]$^{1/2}$)*, which is typical for nanocomposites [26][2]. We will estimate below magnitude of the inter-droplet distance which is becoming large (~100 nm) at the presence of the impurity band. Of course, the model suggested is very simplified and drops of a smaller size and with a smaller hole concentration could exist also. The local ferromagnetic transition in such drops could occur at *T* < $T_c$ ≈ 40 K. That is most probable reason for the weak temperature dependence of *T₀* at those temperatures, which results in a small deviation of experimental curve from the straight line as it is seen in the insert in Fig.1 for *T* < 40 K.

According the above model hole drops are responsible for the Hall effect while compensated regions determine the sample resistance. It is natural because compensated regions where conductivity is of the hopping type does not contribute to the Hall effect and give the main contribution top the sample resistance being mostly resistive. The local ferromagnetic transition results in lowering of the carrier energy inside drops as it is shown by dashed line in Fig. 4 similarly to the formation the bound magnetic polaron (see ref. [28] and references there), and in accordance with recently proposed model [29] (for a-GdSi alloy), the magnetic transition should be accompanied by the increase of the drop size. The simplified physical reason for that could be seen from comparison of dashed and solid line in Fig. 4. The enlarge of the drop size leads to diminishing of inter-drop distance which determines the sample resistance and so $\rho_{xx}$(T) should goes down or at list start to be flatter just after Tc. That explains the appearance of the local maximum or flattening area in the temperature dependence of resistance, as it is observed experimentally (see Fig. 1).

Due to spin-dependent scattering which is reason for NMR the hole mobility inside droplets should be higher in ferromagnetic state than in paramagnetic one [2,3]. So after the local ferromagnetic transition the conductivity of drops increases. Besides, as it was mentioned, similarly to the formation of bound magnetic polaron the energy of drops should fall down that, in its turn, can lead to increase of the holes concentration in ferromagnetic regions (see Fig. 4). Both effects, apparently, could be responsible for the excess of the AHE resistance in paramagnetic state in high magnetic fields (*T* = 88 K) compared it's value in ferromagnetic state (*T* = 25 K, fig. 2) because the AHE coefficient $R_s \propto (\rho_{xx})^\alpha$. Probably, the same reasons are responsible for higher value of the hole concentrations, derived from the $R_H(B)$ slopes at *T* = 25 K in strong magnetic fields, in comparison with that, obtained at *T* = 300 K. It is not possible to exclude, however, that the over-barrier transport of holes at room temperatures (see Fig. 4) could be the reason for underestimation of the holes concentration. The strong difference in observed values of fields of saturation for AHE and MR also is related for different mechanisms responsible for them. While the saturation of AHE is due to saturation of magnetization inside droplets, magnetoresistance is determined by the hopping transport in between drops and follows the magnetic field dependence of the hopping conductivity. If the impurity band is formed by magnetic impurities, its width and the density of states NF at the Fermi's level can be substantially controlled by the fluctuations of the exchange interaction and by the formation of bound polarons [28]. Magnetic field under these conditions leads to the suppression of carrier localization (the spin localization) due to alignment of the magnetic moments of impurities along magnetic field. So NF grows, and parameter T0 determining the hopping conductivity diminishes, that is the main reason for the huge NMR in III-Mn-V semiconductors with the activation type of conductivity [28]. Practically complete alignment of the magnetic moments (> 99%) for paramagnetic ions like Mn2+ is reached at $B/T \approx 4$ T/K [30]. That corresponds with $B \approx 17$ T at $T = 4.2$ K and reasonably correlates with the NMR saturation in our case (Fig.3).

---

[2] In amorphous nanocomposites, where the hopping conductivity can play an essential role in insulating matrix, the appreciable deviation from the law "1/2" and transition to the law "1/4" are observed [27].

Finally, it should be noted that strong variation of the density of states at the Fermi's level under the magnetic field action should be accompanied by the change of topology of the current paths in the percolation cluster. That results in so called effects non-coherent mesoscopics and can be used for estimation of the scales of the magneto-electric nonuniformity [23]. The zero field resistance between Hall probes Ra is due not only to their mismatch, but also due to assymetry of the percolation cluster [23]. The rearrangement of the percolation cluster leads to a variation of $R_a = (R_{xy}^+ + R_{xy}^-)/2$, where the transversal resistances $R_{xy}^+$, $R_{xy}^-$ correspond to positive and negative directions of the external magnetic field.

Normalized dependences of the resistance of asymmetry Ra(B) and the longitudinal resistance Rxx(B) versus magnetic field, measured at $T = 25$ K, are presented in Fig.5. It is seen that $R_a(B)$ qualitatively, distinctly differs from the $R_{xx}(B)$, unlike it should be for transverse resistance due to probes mismatch. That proves the non-coherent mesoscopic reason of the $R_a(B)$ dependence and testifies the percolation type of conductivity in studied films related to the model of the sample structure presented above. The observed deviation of the Ra(B) dependence from the $R_{xx}(B)$ curve can be interpreted as the effective mismatch of the Hall probes Δla, its value is determined by the correlation radius of the percolation cluster $L_c$ ($\Delta l_a \sim L_c$) and following [23] could be estimated as 10 μm. It is not surprising, because $L_c$ includes lots of metallic drops and exceeds their sizes and the inter-droplet distance by several orders of magnitude [23,31]. It is possible estimate magnitudes of these nonuniformities (the drop size and the inter-droplet distance) using approach similar to the one applied in doped compensated semiconductors for estimations of the character scale of the fluctuation potential [17]. Using results of ref. [23,31], one obtains that both are about 100 nm in our case.

**CONCLUSION**

Thus, it is possible to conclude that the peculiarities of investigated In1-xMnxAs (x≈0.1) films caused by the large number of the donor defects and by non-uniform distribution of the active Mn impurities (acceptors), arising after the pulsed laser annealing. As result, the strongly degenerated areas (droplets) with the increased holes concentration is appeared in this system, separated by the compensated spacers, where the holes transport has the hopping character. In accordance with the model of two-component systems [22], the Hall effect is defined by the holes transport through drops, whereas the sample conductivity is determined by the hopping transport in compensated areas. The local ferromagnetic transition which preferably occurs in conductive droplets with enhanced concentration of both magnetic ions and holes leads to a number of new peculiar features of the transport properties in comparison with ordinary two-component systems. The observed anomalies of the Hall effect are caused, to our mind, by the presence of the local ferromagnetic transition accompanied by the lowering of the drops energy and by the increase of their sizes. The decrease of resistance inside droplets causes the surprising behavior of the anomalous Hall effect which in paramagnetic state exceeds it's saturated value below Curie temperature. On the other hand, peculiarities of the negative magnetoresistance are caused by the polaron character of the hopping transport of holes across the impurity band, which is formed by paramagnetic Mn ions under conditions of strong fluctuations of the exchange interaction. The percolation character of conductivity of $In_{1-x}Mn_xAs$ films is proved by the existance of mesoscopic fluctuations of the non-diagonal components of the magnetoresistivity tensor, which analysis gives magnitudes of local nonuniformities about 100 nm.

**ACKNOWLEDGMENTS**

This work is supported by the EU Program EUROMAGNET (contract JRA-IR), the Russian Foundation for Basic Research (projects 04-02-16158 and 05-02-17021) and the ISTC project G-1335.

**Figure Captions**

Fig.1. Curve of specific resistance versus temperature of a InMnAs film. The inserts show the same dependence in the coordinates "$\ln\rho_{xx}$ - $(1/T)^{1/4}$".

Fig.2. The Hall resistance versus magnetic field dependencies of the InMnAs film at different temperatures: 1 – $T$=25 K, 2 – $T$=40 K, 3 – $T$=88 K. The insert shows the $R_H(B)$ curve at $T$ = 25 K in expanded scale.

Fig.3. Specific resistance versus magnetic field dependencies of a InMnAs film at different temperatures: 1 – $T$=25 K, 2 – $T$=40 K, 3 – $T$=88 K. Insert demonstrates the magnetoresistance curve at $T$=4.2 K.

Fig.4. Schematic representation of the holes transport in InMnAs. Arrow 1 indicates the hopping transport of holes through the impurity band in the strongly compensated areas, arrow 2 shows the over-barrier transfer of holes at the percolation level. Solid line indicates the valence band top $E_v$ in paramagnetic state, and dash-dot line shows its shape under conditions of the local ferromagnetic transition. The shaded regions are the states occupied by holes.

Fig.5. Normalized curves describing the magnetic field dependencies of the asymmetry resistance $R_a(B) = [R_{xy}(B) + R_{xy}(-B)]/2$ at $T$ = 25 K (curve 1) and $T$ = 40 K (curve 2) together with the longitudinal resistance $R_{xx}(B)$ at $T$ = 25 K (curve 3).

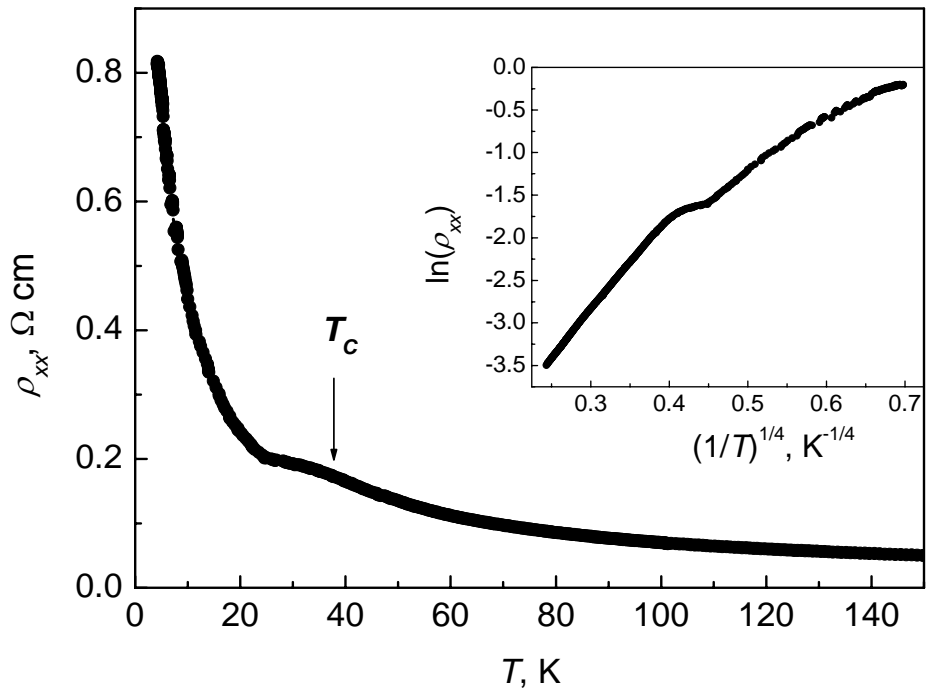

Fig.1

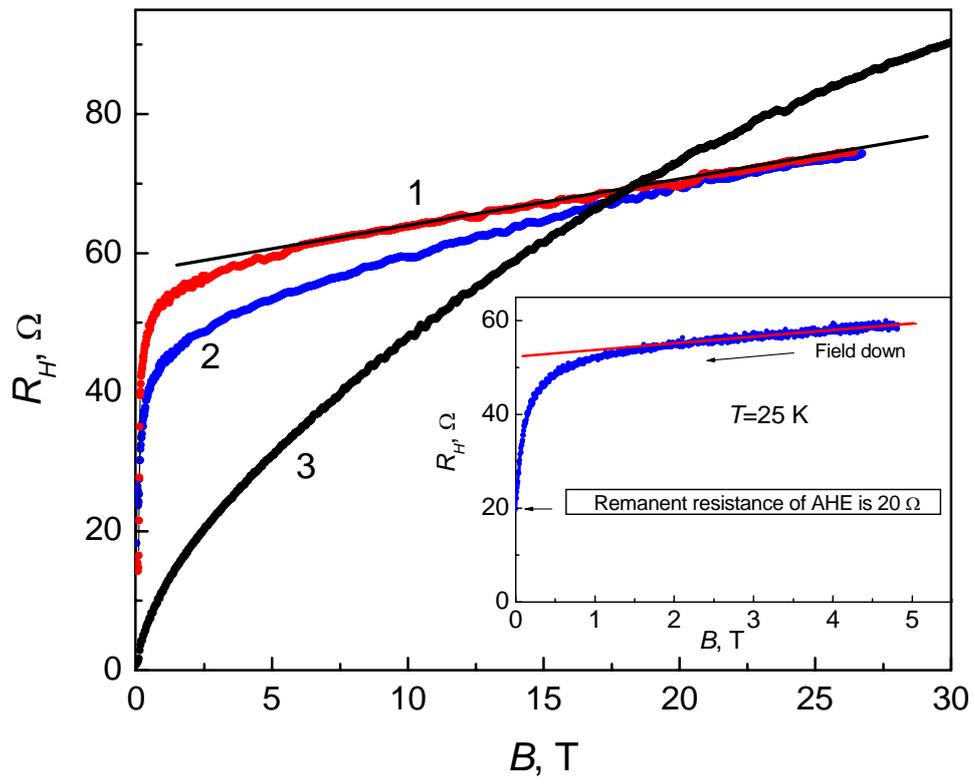

Fig.2

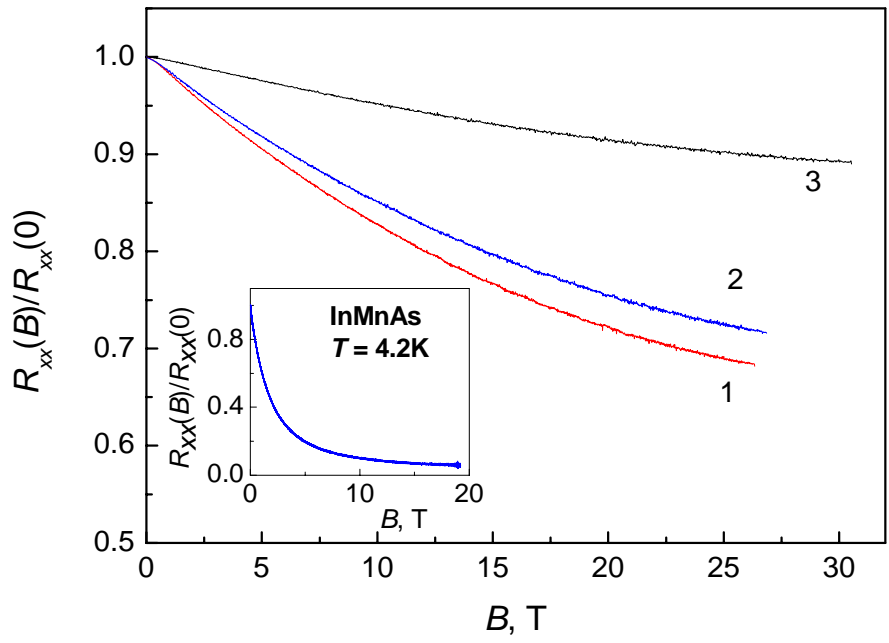

Fig.3

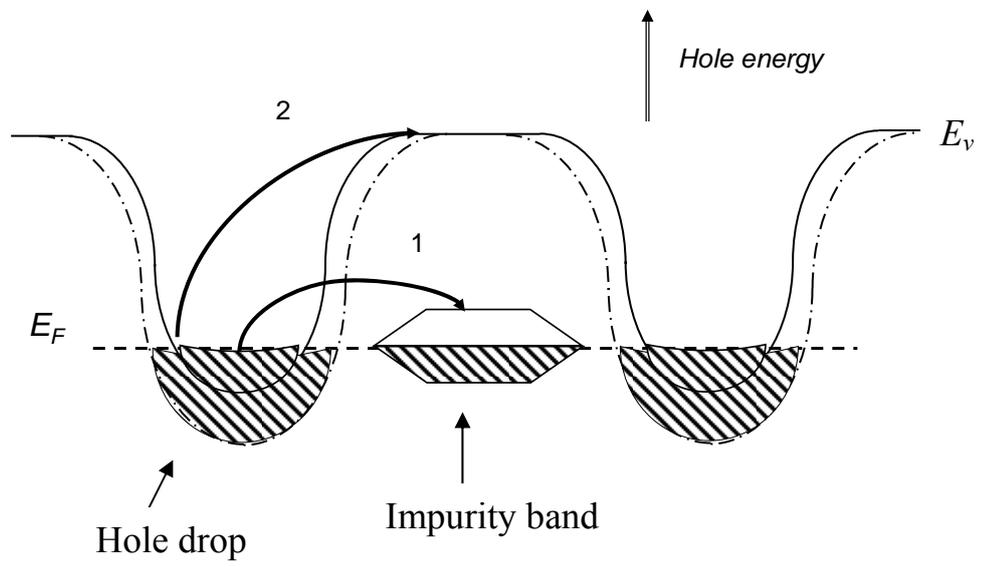

Fig.4

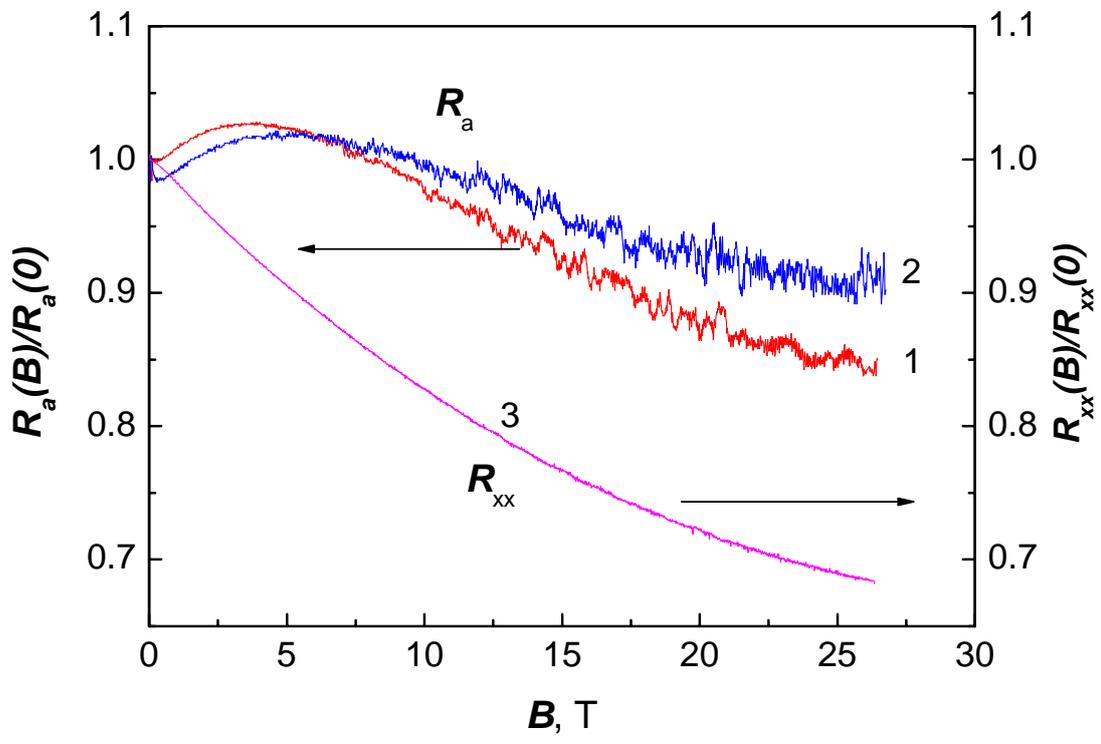

Fig.5